\documentclass[a4paper,prl,superscriptaddress,showpacs,twocolumn]{revtex4}
\usepackage[dvips]{graphicx}
\usepackage{amssymb}
\newcommand{\ped}[1]{\ensuremath{_{\rm #1}}}
\newcommand{\apex}[1]{\ensuremath{^{\rm #1}}}

\begin{document}

\title{Multigap superconductivity and strong electron-boson coupling in Fe-based superconductors:
A point-contact Andreev-reflection study of
Ba(Fe\ped{1-x}Co\ped{x})\ped2As\ped2 single crystals}
\author{M. Tortello}\author{D. Daghero}\author{G.A. Ummarino}
\affiliation{Dipartimento di Fisica, Politecnico di Torino, 10129
Torino, Italy}
\author{V.A. Stepanov}
\affiliation{P.N. Lebedev Physical Institute, Russian Academy of
Sciences, 119991 Moscow, Russia}
\author{J. Jiang}\author{J.D. Weiss}\author{E.E. Hellstrom}
\affiliation{Applied Superconductivity Center, National High
Magnetic Field Laboratory, Florida State University, Tallahassee,
FL 32310, USA}

\author{R.S. Gonnelli \email{E-mail:renato.gonnelli@polito.it}}
\affiliation{Dipartimento di Fisica, Politecnico di Torino, 10129
Torino, Italy}

\begin{abstract}

Directional point-contact Andreev-reflection (PCAR) measurements in
Ba(Fe$\ped{1-x}$Co$\ped{x}$)$\ped2$As$\ped2$ single crystals
($T\ped{c}$=24.5 K) indicate the presence of two superconducting
gaps with no line nodes on the Fermi surface. The PCAR spectra also
feature additional structures related to the electron-boson
interaction, from which the characteristic boson energy $\Omega_b
(T)$ is obtained, very similar to the spin-resonance energy observed
in neutron scattering experiments. Both the gaps and the additional
structures can be reproduced within a three-band $s \pm$ Eliashberg
model by using an electron-boson spectral function peaked at
$\Omega_0=12~\mathrm{meV} \simeq \Omega_b(0)$.
\end{abstract}
 \maketitle

\begin{figure}[t]
\begin{center}
\includegraphics[keepaspectratio, width=1 \columnwidth]{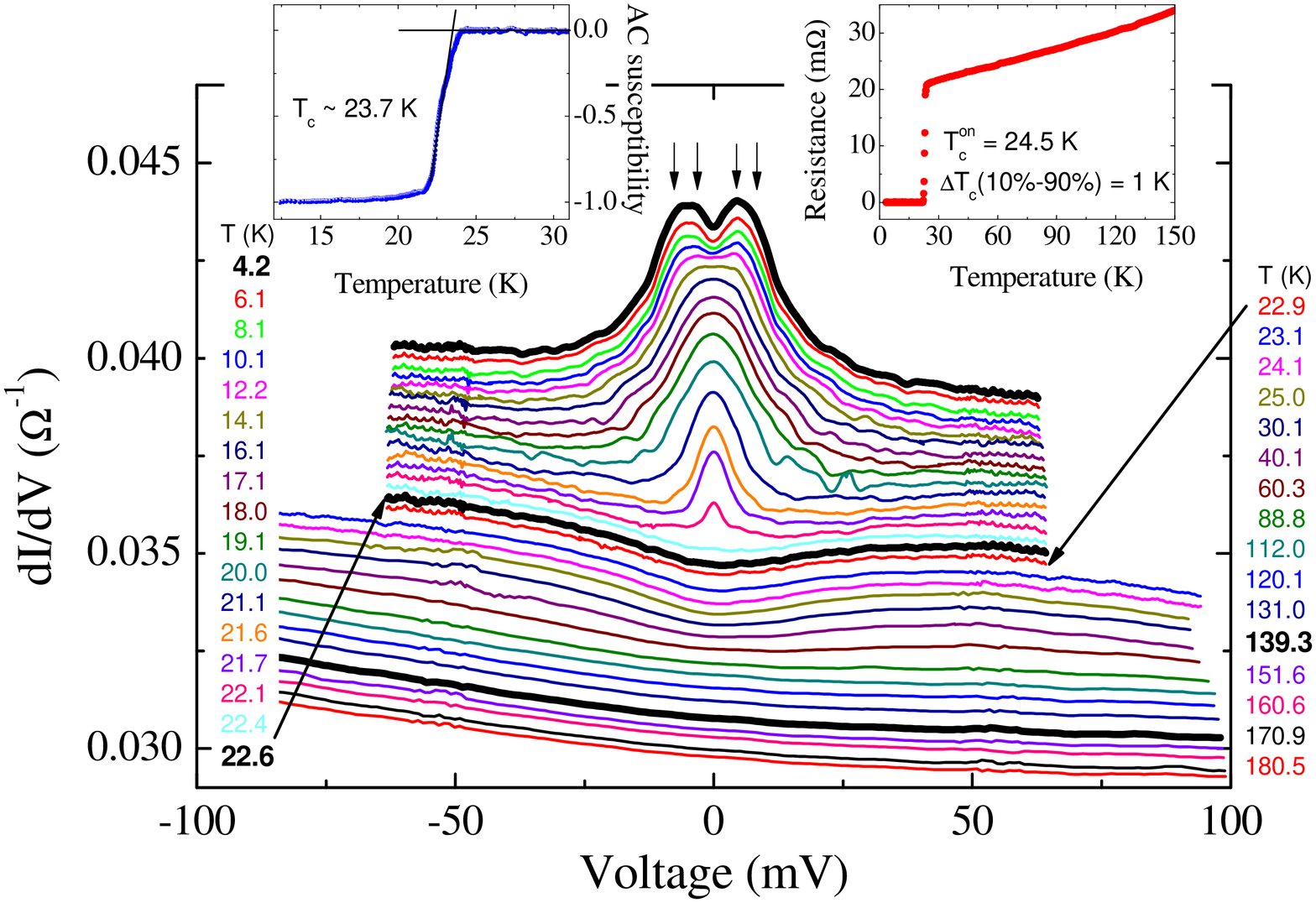}
\end{center}
\vspace{-6mm} \caption{(color online) Temperature dependence of the
differential conductance curves in a
Ag/BaFe$\ped{1.8}$Co$\ped{0.2}$As$\ped2$ $c$-axis point contact. The
curves are vertically offset for clarity. The insets show the
superconducting transition as seen by AC magnetic susceptibility
(left) and DC resistance (right) measurements.}
\vspace{-5mm}\label{Fig1}
\end{figure}

The discovery of the first class of non-cuprate, Fe-based
high-temperature superconductors  in 2008 brought great excitement
in the scientific community \cite{Kamihara_La}. The phase diagram of
these compounds (although still imperfectly known) looks similar to
that of copper-oxide superconductors \cite{Review_Paglione} and, as
in cuprates, superconductivity emerges ``in the vicinity'' of a
magnetic parent compound. The electron-phonon interaction seems not
to be sufficient \cite{Boeri1} to explain their high $T\ped{c}$ (up
to 55 K \cite{Ren_OxDef}) even by considering a magnetic ground
state \cite{Boeri2}. A spin-fluctuation-mediated pairing mechanism
has been early proposed instead, which predicts the occurrence of a
sign change of the order parameter on different sheets of the Fermi
surface ($s \pm$-symmetry) \cite{Mazin_spm}. This picture is
naturally based on the proximity of the superconducting phase to a
magnetic one, on the existence of disconnected Fermi surface (FS)
sheets, and on the multiband character of superconductivity in these
compounds, which are nowadays almost universally accepted
\cite{Mazin_Nature}. The $s \pm$ model itself is strongly supported
by various experimental results \cite{Nat_Phys_spm} which indicate
the existence of multiple nodeless gaps on different sheets of the
FS, although the possible emergence of gap nodes in some systems,
along certain directions or in particular conditions \cite{Kuroki,
Therm_Cond_BaFeCo} is still debated. The role of spin fluctuations
(SF) in the pairing has also found support in neutron scattering
experiments that have revealed a spin resonance energy which scales
linearly with $T\ped{c}$ \cite{Review_Paglione}. Finally, it has
been recently shown that a multiband $s \pm$ Eliashberg model can
reproduce several experimental quantities (such as gaps, $T_{c}$,
kinks in the band dispersion and effective masses \cite{Umma_3 band,
Benfatto_4 band}) by assuming that the mediating boson has a
characteristic energy similar to the spin-resonance one.\\
In this paper we report on \emph{directional} PCAR measurements on
high-quality single crystals of the e-doped 122 compound
BaFe$_{1.8}$Co$_{0.2}$As$\ped 2$. The results prove the existence of
two superconducting gaps with no line nodes on the FS, and whose
amplitude is almost the same in the $ab$ plane or along the $c$
axis. The PCAR spectra also present structures that can be related
to a strong electron-boson interaction (EBI). The characteristic
energy $\Omega_b$ of the mediating boson extracted from the PCAR
curves decreases with temperature and is very similar to the
resonance energy of the spin excitation spectrum \cite{Inosov}.
Moreover, both the gaps and the additional EBI structures in the
PCAR spectra can be reproduced within an effective three-band $s
\pm$ wave Eliashberg model using a boson energy
$\Omega_0=12~\mathrm{meV}\simeq \Omega_b(0)$. All these results
strongly support a spin-fluctuation-mediated mechanism for
superconductivity in this compound.\\
The BaFe$\ped{1.8}$Co$\ped{0.2}$As$\ped2$ (10\% Co) single crystals
were prepared by the self-flux method \cite{Sefat_BaFeCoAs} under a
pressure of 280 MPa at the National High Magnetic Field Laboratory
in Tallahassee. The typical crystal sizes are $\approx
1\times1\times0.1$ mm$\apex3$. The onset of the resistive transition
is $T\ped c \apex{on}=24.5$ K with $\Delta T\ped c$ (10\%-90\%) = 1
K (see inset to Fig.\ref{Fig1}). Instead of using the standard
technique where a sharp metallic tip is pressed against the material
under study, the point contacts were made by putting a small drop of
Ag paste on a fresh surface exposed by breaking the crystal.
Contacts made in this way are very stable and the differential
conductance curves, obtained by numerical differentiation of the I-V
characteristics, can be recorded up to $\approx$ 200 K
\cite{Daghero_Sm}. As an example, Fig.\ref{Fig1} shows the raw
conductance curves, recorded up to 180 K, of a
Ag/BaFe$\ped{1.8}$Co$\ped{0.2}$As$\ped 2$ point contact ($R\ped N =$
25 $\Omega$) with current injection along the $c$ axis (``$c$-axis
contact''). The clear signatures of AR in the low-$T$ curve and the
absence of heating effects or dips \cite{Daghero_review} indicate
ballistic conduction through the point contact, so that
energy-resolved spectroscopy is possible. A closer inspection
reveals that the maxima in the low-$T$ curves present fine
structures (indicated by arrows in Fig.\ref{Fig1}) suggesting
multiple gaps. The Andreev signal decreases on increasing $T$ and
completely disappears at the critical temperature of the contact,
$T\ped{c} \apex{A} =22.6 \pm 0.2$ K, leaving a slightly V-shaped
normal state. On further heating, the normal-state curve
progressively fills and completely flattens at $\approx$ 140 K, the
temperature where the long-range magnetic order sets in in the
parent compound. Similar behavior was observed in 1111 Fe-based
superconductors \cite{LaOFFeAS,Daghero_Sm}.\\
%
%
%
In order to compare the experimental curves to a suitable model, all
the raw conductance curves at $T<T\ped{c}\apex{A}$ were normalized
by the normal-state curve at $T\ped{c}\apex{A}$. Furthermore, to get
rid of the well-known asymmetry of the PCAR spectra of Fe-based
compounds \cite{Daghero_Sm, LaOFFeAS, Szabo_BaKFeAs} the normalized
conductance $G(V)$ was symmetrized, i.e.
$G(V)=[G\ped{exp}(V)+G\ped{exp}(-V)]/2$. This preserves and enhances
the structures we are interested in (gaps and EBI). The asymmetry of
the original curve was taken into account as a source of uncertainty
on the gap values. The resulting conductance curves were then fitted
to a two-band BTK model \footnote{A greater number of bands in the
model implies so many free parameters that the fit becomes
meaningless.} taking into account broadening effects and the angular
distribution of the injected current \cite{Daghero_review}. In this
model the normalized conductance is the weighed sum of two BTK terms
$G(V)=w\ped 1 G\ped 1 (V)+(1-w\ped 1)G \ped 2(V)$, where $w\ped 1$
is the weight of contribution 1. Each term $G\ped i$ is described by
a gap value $\Delta \ped i$, a broadening parameter $\Gamma \ped i$
(here mostly due to inelastic scattering in the vicinity of the
contact) and the parameter $Z \ped i$ which accounts for the height
of the barrier at the N/S interface and the Fermi velocity mismatch \cite{Daghero_review}.\\
%
\begin{figure}
\begin{center}
\includegraphics[keepaspectratio, width=1 \columnwidth]{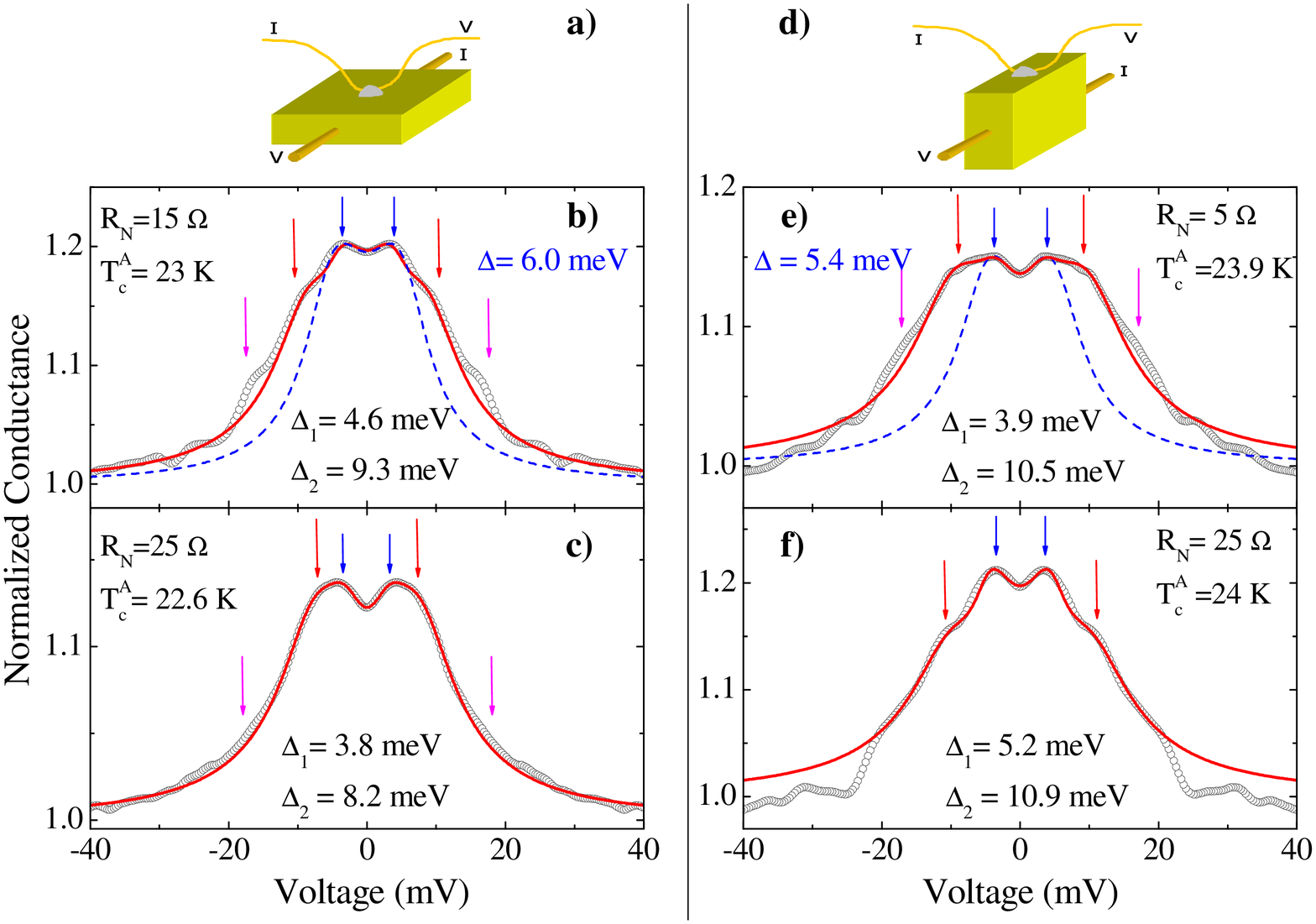}
\end{center}
\vspace{-6mm} \caption{(color online) (a,d) Sketch of $c$-axis and
$ab$-plane contacts. (b,c): normalized conductance curves at 4.2 K
for $c$-axis contacts (symbols) and their two-band fit (solid lines)
with the relevant gap values $\Delta_1$ and $\Delta_2$. Arrows mark
the structures related to the gaps and to the EBI. (e,f): the same
for two $ab$-plane contacts.  In (b) and (e), a single-band fit is
also shown (dashed lines) with the relevant gap amplitude $\Delta$.}
\label{Fig2} \vspace{-5mm}
\end{figure}
%
Fig. \ref{Fig2} shows the setup for PCAR measurements with current
injection along the $c$ axis (a) and along the $ab$ plane (d).
Examples of normalized conductance curves at 4.2 K are shown in (b)
and (c) for $c$-axis contacts and in (e) and (f) for $ab$-plane
contacts. All the PCAR spectra  show peaks at $\approx$ 4 meV and
shoulders at $\approx$ 9-10 meV. Additional structures  are
reproducibly present at 18-20 mV, although more pronounced when the
Andreev signal is higher. In few cases (panel f) they are masked by
small dips, which however do not affect the very clear two-gap
structures at lower energy.
%
Fig. \ref{Fig2} (b) and (e) (as well as the inset to Fig.\ref{Fig3})
clearly show that a one-gap BTK model (dashed line) is unsuited to
reproduce the experimental data while a two-gap model allows a good
fit of the experimental curves, apart from the structures around 20
mV. The resulting amplitudes of the gaps $\Delta\ped 1$ and
$\Delta\ped 2$ are indicated in the labels. In all the two-gap fits
of this paper $w\ped{1} = 0.5 \pm 0.1$ and, at low $T$,
$\Gamma/\Delta = 0.5-0.7$. Finally, $Z$ and $w\ped 1$ are constant
with temperature while $\Gamma$ is almost constant or
slightly increases with $T$ \cite{Daghero_Sm, Daghero_review}.\\
From the two-gap fits of various curves we obtained the average
values: $\Delta \ped 1 \apex c = 4.1 \pm 0.4$ meV and $\Delta \ped 2
\apex c = 9.2 \pm 1.0$ meV for $c$-axis spectra and $\Delta \ped 1
\apex{ab} = 4.4 \pm 0.6$ meV and $\Delta \ped 2 \apex {ab} = 9.9 \pm
1.2$ meV for $ab$-plane contacts. These results can be compared to
ARPES experiments \cite{Ding_PNAS}, which show two nodeless gaps in
the $k\ped{x}k \ped{y}$ plane. The small gap, located on one of the
electron FS sheets, is in very good agreement with our $\Delta\ped
1$. Our value of $\Delta \ped 2$ is instead about 30\% bigger than
the large ARPES gap, located on the hole FS sheet. The reason of
this discrepancy will become clear in the following. In this
concern, note  that, although directional PCAR measurements are not
$k$-resolved, they allow probing the gaps also along the
$\mathbf{k}\ped z$ direction, not easily accessible to ARPES
measurements.\\
The absence of zero-bias conductance peaks (ZBCP) along either
direction in the PCAR spectra rules out line nodes on the FS both
along the $c$ axis and in the $ab$ planes, but does not exclude deep
gap minima or even zeros in small regions of the Brillouin zone
\cite{Therm_Cond_BaFeCo,Mazin_new,Muschler}. The fact that $w\ped1$
is almost independent of the direction suggests an almost equal
degree of three-dimensionality of the various FS sheets in
Ba(Fe$\ped{1-x}$Co$\ped{x}$)$\ped2$As$\ped2$, as also shown by ARPES
\cite{Vilmercati_BaFeCo_3D}, X-ray Compton scattering
\cite{Utfeld} and first-principle calculations \cite{Mazin C,Mazin_new}.\\
Fig. \ref{Fig3} shows the temperature dependence of the normalized
conductance of Fig. \ref{Fig2}(c) (symbols) and the relevant
two-band BTK fit (lines). The two-band model fits very well the PCAR
spectrum at low $T$ (see left inset) giving $\Delta\ped
1(0)\!=\!3.8$ meV and $\Delta\ped 2(0)\!=\!8.2$ meV, which
correspond to $2\Delta\ped 1/k\ped B T\ped c \!\approx \!3.9$ and
$2\Delta\ped 2/k\ped B T\ped c \!\approx \!8.5$, both above the BCS
weak coupling ratio. The temperature dependence of the gaps is
shown in the right inset (symbols).\\
It has been recently shown that in La-1111, Sm-1111 and
Ba$\ped{1-x}$K$\ped{x}$Fe$\ped2$As$\ped2$ the experimental gap
values and their temperature dependence can be reproduced within a
three-band $s \pm$ Eliashberg model \cite{Umma_3 band, Benfatto_4
band}, while two- or three-band weak-coupling BCS models cannot do
the same. In Ba(Fe$\ped{0.9}$Co$\ped{0.1}$)$\ped2$As$\ped2$ we can
simplify the electronic structure, according to ARPES measurements
\cite{Ding_PNAS}, by taking one effective hole band (band 1) and two
electron ones (band 2 and 3, corresponding to the outer and inner
electron barrels in the FS as defined in Ref.\cite{Mazin_new}). We
disregard the small hole pocket at $\Gamma$, predicted by
calculations but not observed by ARPES. Phonons mainly provide
intraband coupling but their contribution is expected to be small
\cite{Boeri1,Boeri2}, while spin fluctuations (SF) mainly provide
the interband coupling. We thus set $\lambda\ped{ii}\apex{ph} = 0.2$
\cite{Boeri1} and
$\lambda\ped{ii}\apex{sf}=\lambda\ped{ij}\apex{ph}=0$ so that the
electron-boson coupling matrix becomes:
\begin{displaymath} \left (
\begin{array}{ccc}
  \lambda \apex{ph} & \lambda_{12} & \lambda_{13} \\
  \lambda_{12}\nu_{12} & \lambda \apex{ph} & 0 \\
  \lambda_{13}\nu_{13} & 0 & \lambda \apex{ph} \\
\end{array}
\right )
\end{displaymath}
where $\nu_{12}=N_{1}(0)/N_{2}(0)$, $\nu_{13}=N_{1}(0)/N_{3}(0)$.
$N_{i}(0)$ is the normal density of states (DOS) at the Fermi level
for the $i$th-band, calculated from the first-principle LDA bands of
the 8\% Co-doped compound \cite{Mazin4}, first shifted downward in
energy and then renormalized by a factor 2 to agree with the ARPES
results \cite{Ding_PNAS, YiPRB}. To satisfy the conservation of the
total charge, the energy shift is 30 meV for the h-bands and 46 meV
for the e-bands. Finally, the total DOS of electron bands is divided
in a 4:1 proportion between bands 2 and 3. This is consistent with
the Raman data \cite{Muschler} that suggest the existence of ``hot
spots'' (where the gap is substantially suppressed) which occupy,
crudely speaking, about 1/2 or less of one out of two electron
pockets \cite{Mazin_new}. This uneven splitting of the DOS is very
important to obtain a satisfactory agreement between the
experimental data and the results of the Eliashberg model. Following
the above, $\nu_{12}= 1.12$ and $\nu_{13}= 4.50$. As for the
electron-SF spectral function, we used a Lorentzian curve peaked at
$\Omega\ped{ij}\!=\!\Omega\ped0\!=\!12$ meV, in agreement with
neutron scattering experiments \cite{Review_Paglione}.\\
%
\begin{figure}[t]
\begin{center}
\includegraphics[keepaspectratio, width=0.9\columnwidth]{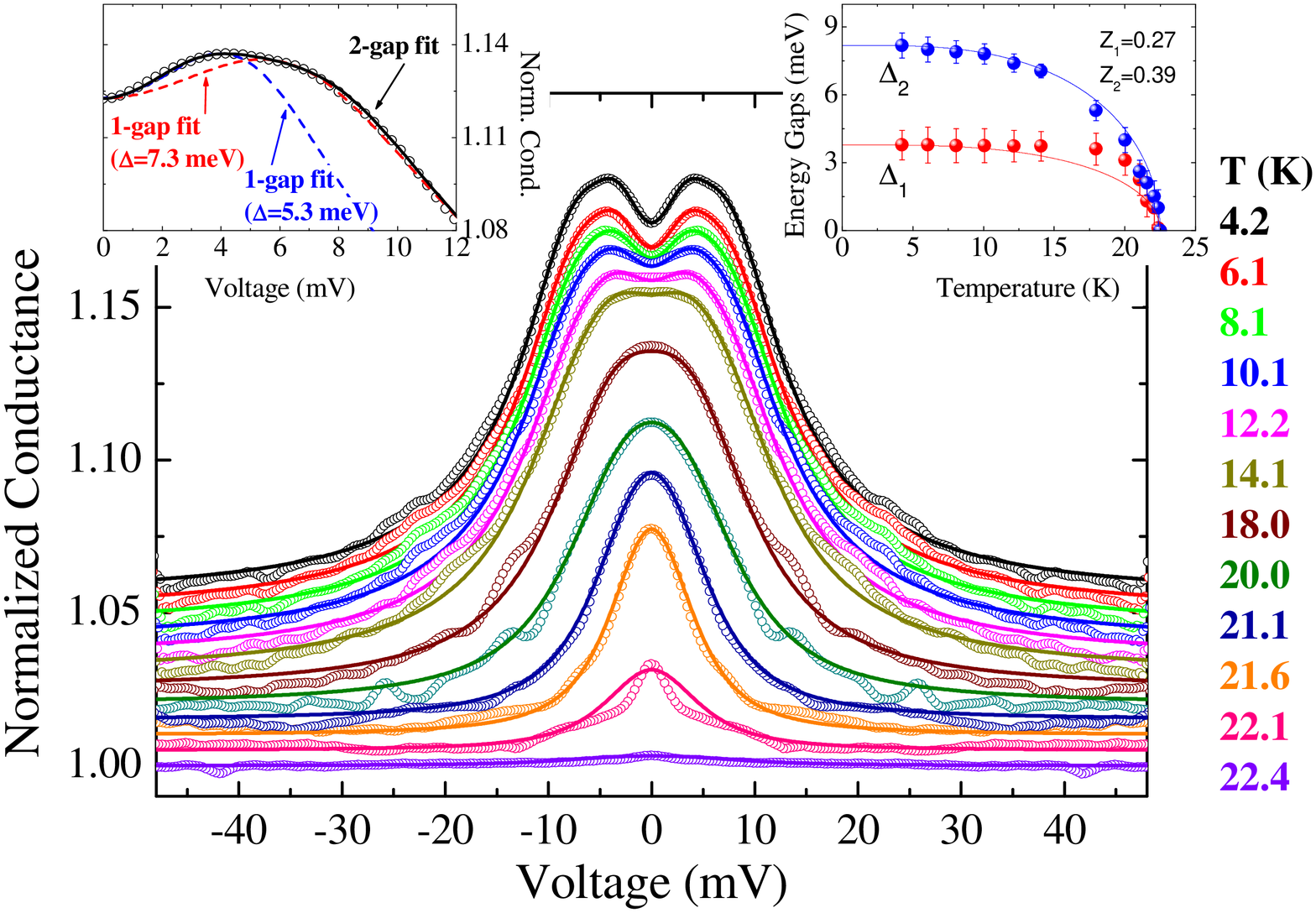}
\end{center}
\vspace{-6mm} %
\caption{(color online) Temperature dependence of the normalized
conductance of Fig.\ref{Fig2}(c) (symbols) and the relevant two-band
BTK fits (lines). All curves except the bottom one are vertically
offset for clarity. The corresponding gaps are shown in the right
inset (symbols) compared with the BCS-like temperature dependencies.
Left inset: zoom of the curve at 4.2 K (symbols) with two possible
one-gap BTK fits (dashed lines) and the best two-band BTK fit (solid
line).} \label{Fig3} \vspace{-5mm}
\end{figure}
%
The only two free parameters of the model are $\lambda_{12}$ and
$\lambda_{13}$ which are chosen so as to reproduce the experimental
gaps as well as possible \cite{Umma_3 band}. The obtained gap values
are $\Delta\ped 1 = 6.1$ meV, $\Delta\ped{2}= -3.8$ meV and
$\Delta\ped{3}= -8.0$ meV (with a theoretical $T\ped c \approx 29.7$
K). $\Delta\ped 1$ (hole FS) and $\Delta\ped 2$ (outer electron FS)
are in very good agreement with the ARPES experiments
\cite{Ding_PNAS}, which actually measured the gap only on one of the
two electron FS sheets. Also, $\Delta\ped 2$ and $\Delta\ped 3$ are
consistent with the gap values observed in our PCAR experiments;
resolving the intermediate gap by PCAR is a challenging task. Thus,
the whole set of data from ARPES, PCAR and calculations looks
consistent. The coupling constants are $\lambda\ped{12}=0.61$ and
$\lambda\ped{13}=1.22$ corresponding to a total effective coupling
constant $\lambda\ped{eff}=1.93$, which indicates,
as expected, a strong-coupling character for this compound.\\
\begin{figure}[t]
\begin{center}
\includegraphics[keepaspectratio, width=0.9\columnwidth]{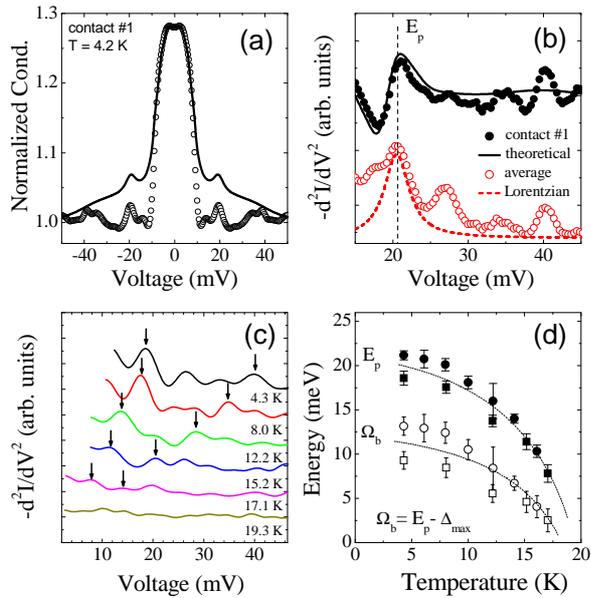}
\end{center}
\vspace{-5mm} %
\caption{(color online) (a) Comparison between an experimental AR
spectrum (symbols) and the theoretical one (line) obtained from
Eliashberg and BTK calculations (see text). (b) Experimental (full
symbols) and theoretical (solid line) $-d^2 I/d V^2$ vs. $V$ curves
obtained from the data in panel (a). Open symbols: the $-d\apex
2I/dV\apex 2$ curve averaged over 5 contacts. Dashed line: the
electron-boson spectral function (shifted in energy by $\approx
\Delta \ped{max}$) used in the three-band Eliashberg calculations.
(c) Evolution of the $-d^2 I/d V^2$ vs. $V$ curves with temperature
showing the displacement of the bosonic structures. The energy of
the peak $E_{p}(T)$ and the corresponding characteristic boson
energy $\Omega_{b}(T)$ are shown in panel (d). Lines are only guides
to the eye.} \label{Fig4} \vspace{-5mm}
\end{figure}
%
Let us now discuss the aforementioned additional structures at about
20 mV that are reproducibly observed in the PCAR spectra (see
fig.\ref{Fig2}), and that disappear at the critical temperature of
the contacts. We will show here that these structures are the
signature of the strong electron-boson coupling, where the boson
characteristic energy is the spin-resonance energy observed by
neutron scattering. Figure \ref{Fig4}(a) shows the normalized
conductance at 4.2 K of a $ab$-plane contact where the AR signal is
particularly high ($ \approx 30\%$), and the structures at $\sim$ 20
mV are clearer than usual, which makes this curve particularly
interesting for our discussion. The solid line is the theoretical
PCAR spectrum obtained from a three-band BTK model by replacing the
constant BCS gaps with the energy-dependent gap functions (for
details on this procedure see $\S$ 4.3.5 of Ref.
\onlinecite{Daghero_review}) calculated within the same Eliashberg
model and with the same parameters discussed above. In the absence
of a theoretical way to account for the broadening parameter
$\Gamma$ within the Eliashberg theory, the diffusive normal
metal/superconductor junction model was used to adjust the amplitude
of the curve to the experimental one \cite{Shigeta_NDiff_S} without
changing the position or shape of its features. This requires fixing
a single parameter $R_d/R_b=1.015$ where $R_d$ ($R_b$) is the
resistance of the diffusive bank (of the junction). The theoretical
AR spectrum clearly shows high-energy structures very similar, in
position and in amplitude, to the experimental ones.\\
Fig. \ref{Fig4}(b) reports the $- d\apex2 I/ d V\apex 2$ curve for
the experimental (full symbols) and theoretical (solid line)
conductance curves shown in figure \ref{Fig4}(a). In
low-transparency (large $Z$) point contacts on strong-coupling
superconductors, peaks in $-d\apex2 I/ d V\apex 2$ correspond to
peaks in the electron-boson spectral function. In the case of small
$Z$, a small relative shift is observed \cite{Daghero_review}, but
here it turns out to be negligible ($< 0.2$ meV). A peak in the
experimental $- d\apex2 I/ d V\apex 2$ is clearly visible at about
21 meV (and is observed also in the theoretical curve). Other
structures appear around 27 mV and 40 mV. All these structures exist
also in the $-d\apex2 I/ d V\apex 2$ curve obtained by averaging
over 5 different contacts (open symbols). The energy of the first
maximum, $E_p$, agrees well with the energy of the peak in the
Lorentzian electron-boson spectrum used in our calculations, shifted
by $\sim \Delta \ped{max}$ (dashed line) \cite{Daghero_review},
further indicating that a bosonic mode at $\Omega_0$ is really
playing a major role in the coupling. The structures at higher
voltage that do not appear in the theoretical $-d\apex2 I/ d V\apex
2$ (solid line in Fig. \ref{Fig4}b) may be due to the actual shape
of the electron-SF spectral function and/or to non-linear
strong-coupling effects. Fig.\ref{Fig4}(c) shows that, on increasing
temperature, all the EBI structures shift to lower energy.
Fig.\ref{Fig4}(d) reports the maximum and minimum values of $E_{p}$
over the different $-d\apex2 I/ d V\apex 2$ curves (full symbols),
and of the quantity $E_{p}-\Delta_{max}$ (open symbols) as a
function of temperature. Note that the latter is the energy of the
``resonant mode'' in the electron-boson spectrum, $\Omega_{b}$
($\Omega_{b}\simeq \Omega_0$ at low $T$) \cite{Popovich} and its
behavior is indeed very similar to that of the spin resonance energy
measured by neutron scattering experiments \cite{Inosov}.\\
In conclusion, we have shown that PCAR measurements give direct and
clear evidence for multiband strong coupling superconductivity in
Ba(Fe$\ped{1-x}$Co$\ped{x}$)$\ped 2$As$\ped2$. They also allow
extracting the characteristic energy of the mediating boson and its
$T$ dependence, that both coincide with those of the spin resonance
measured by neutron scattering experiments \cite{Inosov}. This
brings unambiguous evidence for a  spin-fluctuation-mediated $s \pm$
mechanism of superconductivity in this compound.\\
We wish to thank M. Putti for providing the samples, I.I. Mazin and
E. Cappelluti for invaluable discussions. This work was done within
the PRIN project No. 2008XWLWF9-005.

\end{document}